\documentclass[12pt] {article}
\usepackage{amsmath,bm}

\renewcommand{\baselinestretch}{1.68}

\usepackage{empheq}

\usepackage{graphicx}
\usepackage{color}

\usepackage{lscape, comment, latexsym, array, amssymb, fullpage}

\newcommand{\sig}{\sigma}

\newcommand{\bth}{\bm{\theta}}

\newcommand{\bmu}{\bm{\mu}}
\newcommand{\bbeta}{\bm{\beta}}

\newcommand{\bgamma}{\bm{\gamma}}
\newcommand{\bkpa}{\bm{\kappa}}

\newcommand{\bmeta}{\bm{\eta}}

\newcommand{\bX}{\bm{X}}
\newcommand{\bx}{\bm{x}}

\newcommand{\by}{\bm{y}}
\newcommand{\bu}{\bm{u}}

\newcommand{\bz}{\bm{z}}
\newcommand{\bD}{\bm{D}}

\newcommand{\bW}{\bm{W}}

\newcommand{\cA}{{\cal A}}

\newcommand{\ty}{\tilde{y}}
\newcommand{\tz}{\tilde{z}}

\newcommand{\tby}{\tilde{{\bm y}}}
\newcommand{\tbx}{\tilde{{\bm x}}}

\newcommand{\PT}{\mbox{PT}}

\newcommand{\deltabar}{\overline\delta}

\newcommand{\Wxj}{W_j}

\newcommand{\PCTETTE}{\mbox{PCTE$^{\rm{\, TTE}}$}}
\newcommand{\PCTETox}{\mbox{PCTE$^{\rm{\, tox}}$}}

\newcommand{\PATESSTTE}{\mbox{PATE$_{SS}^{\rm{\, TTE}}$}}
\newcommand{\PATESSTox}{\mbox{PATE$_{SS}^{\rm{\, tox}}$}}

\newcommand{\SG}{\mbox{SG}}
\renewcommand{\Pr}{\mbox{Pr}}

\newcommand{\ul}[1]{\underline{#1}}
\newcommand{\ytox}{y^{\rm\, tox}}
\newcommand{\bytox}{\bm{y}^{\rm \, tox}}
\newcommand{\tytox}{{\tilde{y}}^{\rm\, tox}}
\newcommand{\bmutox}{\bm{\mu}^{\rm\, tox}}

\newcommand{\ptox}{p^{0}_{\mbox{tox}}} 
\newcommand{\bxtox}{\bm{x}^{\mbox{tox}}}
\newcommand{\bthtox}{\bm{\theta}^{0,\mbox{tox}}}

\newcommand{\bbetatox}{\bm{\beta}^{0,\mbox{tox}}}
\newcommand{\gammatox}{\gamma^{0,\mbox{tox}}}
\newcommand{\gtox}{g_{\mbox{tox}}} 
\newcommand{\htox}{h_{\mbox{tox}}} 

\newcommand{\DTTE}{\bm{D}^{\rm{TTE}}}  
\newcommand{\Dtox}{\bm{D}^{\mbox{tox}}} 
\newcommand{\obs}{^{\rm obs}}
\newcommand{\unobs}{^{\rm unobs}}
\newcommand{\tte}{^{\rm TTE}}
\newcommand{\tox}{^{\rm tox}}

\begin{document}

\thispagestyle{empty}

\centerline{\large \bf A Semi-parametric Bayesian Approach to Population Finding with}  
\centerline{\large \bf Time-to-Event and Toxicity Data in a Randomized Clinical Trial}

\vskip .3in 

\centerline{Satoshi Morita$^{1*}$, Peter M\"uller$^{2}$, and Hiroyasu Abe$^{1}$}

\vskip .1in
\centerline{$^{1}$Department of Biomedical Statistics and Bioinformatics,}
\centerline{Kyoto University Graduate School of Medicine, Kyoto, Japan}
\centerline{$^{2}$Department of Mathematics, University of Texas, Austin, TX, U.S.A.}

\centerline{$^*${\it e-mail}: smorita@kuhp.kyoto-u.ac.jp}

\vskip .2in

\noindent {\sc Summary}.   
A utility-based Bayesian population finding (BaPoFi) method was
proposed by Morita and M\"uller (2017, {\it Biometrics}, 1355-1365) to
analyze data from a randomized clinical trial with the aim of
identifying good predictive baseline covariates for optimizing the
target population for a future study. 
The approach casts the
population finding process as a formal decision problem together with
a flexible probability model using a random forest to define a
regression mean function.  BaPoFi is constructed to handle a single
continuous or binary outcome variable.  In this paper, we develop
BaPoFi-TTE as an extension of the earlier approach for clinically
important cases of time-to-event (TTE) data with censoring, 
and also accounting for a toxicity outcome. 
We model the
association of TTE data with baseline covariates using a
semi-parametric failure time model with a P\'olya tree prior for an
unknown error term and a random forest for a flexible regression mean
function.  We define a utility function that addresses a trade-off
between efficacy and toxicity as one of the important clinical
considerations for population finding.  We examine the operating
characteristics of the proposed method in extensive simulation
studies.  For illustration, we apply the proposed method to data from
a randomized oncology clinical trial.
Concerns in a preliminary analysis of the same data based on a
parametric model motivated the proposed more general approach. 


\vskip .20in
\noindent {\sc Short title}:  
Semi-parametric Bayesian population finding.

\vskip .3in
\noindent $^*$ Address for correspondence:

Satoshi Morita, PhD

Department of Biomedical Statistics and Bioinformatics, 

Kyoto University Graduate School of Medicine

Address: 54 Kawahara-cho, Shogoin, Sakyo-ku, Kyoto 606-8507, Japan

E-mail: smorita@kuhp.kyoto-u.ac.jp

Phone: +81-75-751-4717, Fax: +81-75-751-4767

\thispagestyle{empty}

\newpage
\section{Introduction}
\label{sec:Introduction}

\setcounter{page}{1}

\noindent 
We consider finding sensitive subpopulations for a new treatment based 
on data of time-to-event (TTE) efficacy and binary toxicity outcomes in a randomized
clinical trial (RCT).
Such approaches are needed, for example, for oncology RCTs that usually evaluate
clinical benefit for patients using TTE outcomes, such as
progression-free survival (PFS) and/or overall survival (OS) as the
primary endpoint(s), and also account for toxicity.
Investigators are interested in identifying proper predictive baseline
covariates including clinical characteristics and biomarkers of
patients to optimize the target population for further treatment
development, and for treatment individualization.  The importance of
examining predictive covariates for efficacy and toxicity of a new
treatment in a clinical development has been extensively discussed,
among others, by 
Renfro {\it et al.} (2016) and Ondra {\it et al.} (2016), etc.  

We approach the problem as a Bayesian decision problem.  
Methodologically, we separate the construction of the 
statistical inference model to fit the data, and a description of the decision, including possible actions and preferences,
building on Sivaganesan {\it et al.} (2017)  and Morita and M\"uller (2017),
who propose
a utility-based Bayesian population finding (BaPoFi) method that casts
the population finding process as a formal decision problem.  The
approach is valid with any sufficiently flexible data model, including
Bayesian additive regression trees (BART) (Chipman {\it et al.}, 2010, Hill, 2011), 
multivariate adaptive regression splines (MARS) (Friedman, 1991),
classification and regression trees (CART) (Chipman {\it et al.},
1998),  or
Gaussian processes (Rasmussen and Williams, 2006).
BaPoFi uses BART to evaluate enhanced treatment effects
based on a continuous or binary outcome in patient subpopulations
defined by baseline covariates in an RCT.
However, BaPoFi does not work well for TTE data with censoring.  
The utility function embedded in BaPoFi evaluates effect sizes for
efficacy, population sizes, and numbers of covariates 
that are needed  to describe
subpopulations, but does not account for safety aspects of
treatments. 

In this paper, we propose BaPoFi-TTE as a generalization of BaPoFi for TTE outcomes and 
introduce explicit efficacy-toxicity trade-offs to
provide a  practically useful method for
population finding in
the clinical development of a new treatment.
To implement such inference we model both, efficacy and toxicity
outcomes, and define the notion of 
a minimum clinically meaningful difference (MCMD)
in efficacy that is explicitly allowed to vary as a function of
toxicity. 
Another important feature of the proposed approach is the 
use of a semi-parametric Bayesian
accelerated failure time (AFT) model with a random forest implemented
in BART as a regression mean function,
thereby avoiding restrictive parametric assumptions.
For the same reason, for the unknown residual distribution in the
AFT model, we assume a P\'olya tree prior (Hanson, 2006).
An interesting alternative model for similar data is the BART for TTE
probit model for distinct event and censoring times developed by 
Sparapani {\it et al.} (2016).  We use a second instance of BART to model a binary
toxicity outcome.  This allows us to introduce
efficacy-toxicity trade-offs in the utility function. 

There is a substantial literature on methods for patient subpopulation
finding with enhanced treatment effects.  Foster {\it et al.} (2011)
developed a tree-based algorithm to evaluate enhanced treatment
effects in patient subgroups.  Lipkovich {\it et al.} (2011) used a
recursive partitioning method to identify patient subgroups with
different responses to a treatment.  Schnell {\it et al}. (2016)
developed a Bayesian credible subgroups method to identify a benefiting
subgroup for a treatment.  Jones {\it et
al}. (2011) give a good review of Bayesian approaches to subgroup
analysis.  For decision making related to the study design, Simon and
Simon (2018) proposed a framework for group-sequential adaptive
enrichment clinical trials using Bayesian methods
with similar modeling features. 
Xu {\it et al.}
(2018) proposed an adaptive approach for a master protocol clinical
trial design
using a non-parametric Bayesian model and a utility function for adaptive
allocation and subgroup finding, similar to the utility introduced in the
upcoming discussion. 
Graf {\it et al.} (2015) discuss utility functions to evaluate a
benefit risk balance of a treatment in adaptive study designs for
subpopulation analysis.
\smallskip

The rest of this paper is organized as follows.  In Section 2, we
provide the motivating example.  Section 3 presents our proposed
approach to finding a sensitive subpopulation as a decision problem.
We introduce a probability model to summarize TTE data in Section 4, and
briefly describe the posterior computations in Section 5.  In Section 6, we
conduct extensive simulation studies to examine the operating
characteristics of the proposed method.  We present 
results for the motivating study in 
Section 7, and close with a brief discussion in Section 8.

\section{Motivating example}   
\label{sec:MotivaingEx}

\noindent We analyze data from an oncology randomized clinical trial
to find a sensitive subpopulation for a new therapy.  To evaluate
sensitive patient subpopulations, clinical papers usually report 
extensively so-called subgroup analyses.  However,
carrying out subgroup analyses with a large number of
candidate covariates gives rise to multiplicity issues due to multiple
testing, and clinical concerns related to the difficulty and
complexity in interpreting many between-treatment-arm comparisons.
Such concerns are exacerbated when 
clinical investigators are interested in even higher order interaction
effects, e.g., between treatment 
and multiple covariates.

The proposed approach is motivated by an analysis in 
Twelves {\it et al.} (2016) who carried out data analysis
for OS data from a randomized phase III clinical trial in
patients with locally advanced or metastatic breast cancer (Kaufman
{\it et al.}, 2015).  They examined the influence of baseline patient
clinical characteristics on OS.  In this trial, 544 and
546 (in total, 1090) patients received experimental and control
treatments, respectively.  The study evaluated OS and PFS as
co-primary endpoints.  Although a statistically significant difference
in OS was not observed between the two treatments, Kaplan-Meier
curves comparing the two arms suggested an improvement in OS with the
experimental treatment.  For safety evaluation, a noticeable
difference in grade 4 hematologic toxicities was reported.

This motivates us to consider a formal subgroup analysis, to
search for a benefiting subpopulation of patients, allowing for an
efficacy-toxicity tradeoff. We will report details of the analysis
and results later. 
In short, the proposed method finds triple negative status as an
important
baseline covariate to define the benefiting subpopulation, together
with several other covariates.  We 
will formally describe how the size of the reported
subpopulation varies with the efficacy and toxicity trade-off and the
tuning parameters in the utility function.

\section{Population finding}
\label{sec:Popfinding}

\noindent Our approach is based on casting inference for population
finding as a formal decision problem.  The basic components of a
decision-theoretic setup include an action space ${\cA}$ of possible
decisions $a \in \cA$, a probability model $p(\by,\bmu)$ for data
$\by$ and parameters $\bmu$, and a utility function $u(a, \bmu, \by)$.
For the moment we do not need any details of the probability model,
and defer the discussion of a specific model for later, in Section \ref{sec:ProbModel}. 
Data may include observed and future data.  We will use $\tby$ to separately indicate
future data, when needed. 
A utility function $u(\cdot)$ quantifies relative
preferences for hypothetical future outcomes ($\tby$) and assumed
parameter values ($\bmu$) under alternative decisions $a$, given
observed data ($\by$).
It can be argued that a rational decision maker should choose an
action in $\cA$ to maximize $u$ in expectation (Robert, 2007).
The expectation is with respect to $p(\cdot)$, conditioning on all observed
data, and marginalizing over all parameters and future data.  We
will use $U(a)$ to denote expected utility.

\subsection{Notation}
\label{sec:Notation}

\noindent We consider an RCT comparing a TTE outcome, e.g., PFS and OS
time, between control ($C$) and experimental therapy ($N$) arms with a
total sample size of $n$ patients.  Let $T_i$ denote the event time
for patient $i$.
Introducing an event indicator $\gamma_i$, let $T^o_i$ denote either
the observed event time or a (right) censoring time, that is, 
$\gamma_i = 1$ if $T^o_i = T_i$ and $\gamma_i = 0$ if $T^o_i < T_i$.  
Denoting $y_i = log(T_i)$ and $y^o_i = log(T^o_i)$, 
let $(\by^o, \bgamma)$ = $\bigl( (y^o_1,\gamma_1), \ldots, (y^o_n,\gamma_n)\bigr)$
denote TTE  data for all $n$ patients. 
In addition, let $\bytox$ = ($\ytox_1,\ldots,\ytox_n$),
where $\ytox_i = 1$ refers to a toxicity event for patient $i$. 
Let 
$\bz$ = ($z_1,\ldots,z_n$) denote treatment indicators with $z_i = 0$ for arm $C$ and
$z_i = 1$ for arm $N$, and let $\bx_i = (x_{i1},\ldots,x_{ip})$ 
and $\bX = (\bx_1,\ldots,\bx_n)$ denote patient baseline covariates.
Throughout, $\bx, z, y$ without $_i$ index denotes data for a
generic patient. 
In summary, $\bD = (\bz, \bX, \by^o, \bgamma, \bytox)$ denotes
all data observed in $n$ patients in the clinical trial. 
Let $\bmu$ and $\bmutox$ denote parameters that index the
sampling model for the TTE and toxicity outcomes, respectively, given
treatment assignment and covariates. 
In addition, we introduce a nonparametric prior model for the residual
distribution $G$ in the TTE sampling model, with the prior
involving an additional variance parameter $\sig^2$ (details later). 
We separately analyze TTE and toxicity outcomes, thus separately analyzing 
$\DTTE = (\bz, \bX, \by^o, \bgamma)$ and
$\Dtox = (\bz, \bX, \bytox)$, as is 
usually done in RCTs.  

Introducing additional notation, $log(T_i)$ = $y_i\unobs$ =
$y_i^o + \kappa_i$ for censored cases ($\gamma_i=0$),
let $\by\unobs$ and $\bkpa$ denote the vectors of $y_i^{\rm
  unobs}$ 
and $\kappa_i$ for censored cases, respectively, and let
$\by\obs$ denote the vector of observed $y_i^{o}$ for uncensored
cases ($\gamma_i = 1$). 
The observed data $(\by^o, \bgamma)$ together with $\bkpa$ 
define the complete data $\by = (\by\obs, \by\unobs)$.
These imply a joint posterior probability model $p(\bmu, \sigma^2, \bkpa \mid \DTTE)$ 
and a posterior predictive distribution 
$p_{pp} \bigl( \ty_{n+1} \mid z_{n+1}, \bx_{n+1}, \DTTE \bigr)$ 
for a predicted outcome $\ty_{n+1}$ of a future patient $i $= $n$+1.
If it is judged clinically relevant to account for toxicity in
conjunction with efficacy, 
we additionally evaluate a posterior predictive distribution 
$p_{pp}\bigl(\tytox_{n+1} \mid z_{n+1}, \bx_{n+1}, \Dtox\bigr)$ 
of a future toxicity observation, $\tytox_{n+1}$, which is 
derived using a posterior probability model $p(\bmutox \mid \Dtox)$.
Note the notation $p_{pp}(\cdot)$ for the two predictive
distributions. 
In all the above computations, we assume {\it a priori} independent
$\bmu$, $\sigma^2$, and $\bmutox$.

\subsection{Subgroups}
\label{sec:Subgroups}

\noindent 
Our approach includes a characterization of subpopulations on the
basis of discretized and categorical covariates, $x_j$, $j =
1,\ldots,p$.  For continuous covariates, taking typical sample sizes
of clinical trials into account, we consider trichotomizing each
covariate.  Let $Q_j^{33}$ and $Q_j^{67}$ denote the 33\% and 67\%
quantiles of $x_j$.  For categorical covariates with more than three
categories, we consider merging categories to fewer, clinically
meaningful categories.  Let $\{M_1, \ldots, M_d\}$ generically denote
the two ($d=2$) or three ($d=3$) categories of a dichotomized or
trichotomized categorical covariate.  We use subsets $\Wxj \subset
\{M_1, \ldots, M_d\}$ to describe subgroups of patients.

We report a subgroup of most benefiting patients based on
trichotomized (or dichotomized) covariate values $W_j$ as follows
\begin{equation} 
  a = \bigl(J,{\bW} \bigr),
\label{subgpselect}
\end{equation}
\noindent where $J \subseteq \{1,\ldots,p\}$ indicates the covariates
that characterize the subgroup, and $\bW = \bigl\{\Wxj, \ j \in J
\bigr\}$ indicates the levels of those covariates.  We write $\{i:\;
\bx_i \in a \}$ as short for $\{i:\; x_{i,j} \in \Wxj \mbox{ for } j
\in J\}$.  
Let SG$(a) = \{i:\; \bx_i \in a \}$ denote patients within the subgroup
selected by $a$.  For example, we may report that the patient subgroup
$a= \bigl(j, \Wxj = \{M_2,M_3\} \bigr)$ is sensitive to treatment $N$
and $x_j$ is the predictive covariate of the subgroup.  In addition,
we introduce two more special cases, $a$ = \lq\lq{\it null"} for
reporting that treatments $N$ and $C$ show same efficacy and toxicity
effects in any populations, and $a$ = \lq\lq{\it all"} for reporting
that treatment $N$ is more effective than treatment $C$, equally so
for the entire population of patients.

In the present study, we consider only sensitive subgroups that are
defined by one or two predictive covariates.  We introduce this
restriction because three or more predictors may not be interpretable
or practically useful, and also to keep the computational effort
reasonable.  Thus, we restrict to $J = \{j\}$ or $J = \{j, k\}$.  For
the one-covariate cases, $J = \{j\}$, we evaluate all the six subsets:
$W_j = \{M_1\}$, $\{M_2\}$, $\{M_3\}$, $\{M_1,M_2\}$, $\{M_2,M_3\}$,
and $\{M_1,M_3\}$ with respect to every covariate $x_j$, $j =
1,\ldots,p$.  For the two-covariate cases, $J=\{j,k\}$, we construct
all the possible combinations of one from the six subsets of $x_j$ and
one from those of $x_k$, resulting in rectangular subgroups or
$L$-shaped subgroups.  A rectangular subgroup is constructed as
$\{\bx: x_{i,j} \in W_j \mbox{ and } x_{i,k} \in W_k\}$.  $L$-shaped
subgroups are an ad-hoc way of slightly generalizing the rectangular
subgroups to $\{\bx: x_{i,j} \in W_j \mbox{ or } x_{i,k} \in W_k\}$
(by allowing the union instead of the intersection).  The two restrictions on the action
space ${\cA}$ by discretizing covariates and limiting the number of
predictive covariates, contribute to make the proposed method
practically useful, as will be discussed later in the simulation and
application sections.

\subsection{Utility function}
\label{sec:UFnct}

To examine heterogeneous treatment effects depending on patient
covariates, we use a potential outcomes framework, and introduce potential outcomes 
$\bigl\{y_{n+1} (C), y_{n+1} (N)\bigr\}$ and $\bigl\{\ytox_{n+1} (C),
\ytox_{n+1} (N)\bigr\}$. 
respectively. In actual data, of course only one element of the
pair is observed (Rubin, 1978; Hill, 2011).
For a TTE outcome, letting $S(t)$ denote the survival probability
at time $t$, we evaluate the differences in survival
probabilities, $S(\tau \mid z=1, \bx) - S(\tau \mid z=0, \bx)$.
The threshold $\tau$
is chosen to be a meaningful time horizon for the
condition and treatment under consideration.  For a binary toxicity
outcome, we evaluate the differences in toxicity probabilities between
treatment arms.

Recall that $\DTTE$ and $\Dtox$ denote all observed
efficacy and toxicity data. 
In preparation for the upcoming construction of a utility function, we define the predictive
conditional treatment effect (PCTE) for a future patient $i$=$n+1$.
Let $\tbx=\bx_{n+1}$, $\ty=y_{n+1}$, $\tytox=\ytox_{n+1}$, and
$\tz=z_{n+1}$. 
We define 
\begin{equation}
\PCTETTE(\tbx, \DTTE) = 
S(\tau \mid \tz=1,\tbx, \DTTE) - S(\tau \mid \tz=0,\tbx, \DTTE),
\label{PCTETTE}
\end{equation}
where the probabilities are calculated with respect to 
the posterior predictive distribution $p_{pp}(\ty \mid \tz, \tbx, \DTTE)$. Similarly,  
$\PCTETox(\tbx, \Dtox) = E \bigl\{ \tytox(N) - \tytox(C)  \mid \tbx, \Dtox \bigr\}, $
where the expectation is with respect to the posterior predictive
distribution $p_{pp}(\tytox \mid \tz, \tbx, \Dtox)$.
Finally,  we define the predictive average treatment effect (PATE) for
the TTE outcome in a selected subgroup by averaging
$\PCTETTE(\tbx, \DTTE)$ over $\tbx$ in subgroup $a$, as 
\begin{equation}
\PATESSTTE(a) = \frac{1}{\left| \SG(a) \right|} \cdot \sum_{i \in
  \SG(a)} \PCTETTE(\bx_i, \DTTE). 
\label{PATE_SS_AV}
\end{equation}
The sum is over all observed patients in $\SG(a)$, implying the use of
the empirical distribution to average over patients with covariates in 
subgroup $a$. 
We similarly define $\PATESSTox(a)$ for the toxicity outcome.

Next, we define a minimum clinically meaningful difference in efficacy $\delta$ (MCMD).
We allow $\delta$ to vary as a function of $\ytox$, i.e., $\delta
\left\{\ytox (N), \ytox (C) \right\}$, and define an average MCMD over
a subgroup as
\begin{equation}
\deltabar_a = \sum_{i \in \SG(a)}
      E\left( \delta \left\{\tytox (N), \tytox (C)  \mid \tbx=\bx_i, \Dtox \right\} \right).  \\
  \label{deltabar}
\end{equation}
Here, the expectation is with respect to $p_{pp}(\tytox \mid \tbx, \tz, \Dtox)$. 
As a specific function 
$\delta(\cdot)$ we use $\delta = \delta_0 + \delta_1 \bigl\{\ytox (N) -
\ytox (C) \bigr\}$, where $\delta_0$ has the interpretation of an MCMD
under no difference in toxicity between $N$ and $C$ and $\delta_1$ is
a slope.  
In this case, $\deltabar_a$ reduces to $\deltabar_a = \delta_0 +
\delta_1 \PATESSTox(a)$.  

Finally, using PATE and $\deltabar_a$, we then propose a utility
function
to formalize preferences in terms of efficacy and toxicity across
possible 
actions of subgroup reporting ($a$).
\begin{equation}
  U(a) =  
  \begin{cases}
    \left[\PATESSTTE(a) - \deltabar_a \right]\,
    \frac{|SG(a) + 1|^{\nu}}
         {\left(|J| + 1\right)^{\zeta}} &
    \mbox{if } a \not= null\\
    \ \ \ \ \ \ u_0 &
    \mbox{if } a = null
  \end{cases}
  \label{Utility}
\end{equation}
where the positive constants $(\nu, \zeta, u_0)$ are tuning parameters. 
Later, in Section \ref{sec:Freqeval}, we will show how $(\nu,\zeta)$ can be used to achieve desired frequentist operating characteristics.
In addition, $u_0$ specifies the utility for the action $a$ = $null$,
and is a convenient tuning parameter to achieve a desired type I error rate. 
Since PATE and $\deltabar_a$ already include marginalization
w.r.t. $\ty$, the function \eqref{Utility} is already an expected
utility. Writing \eqref{Utility} as a product of three factors,
$$
U(a)=
E\left\{ m \left[a, \bigl\{\ty (N),    \ty (C)\bigr\},
                    \bigl\{\tytox (N), \tytox (C)\bigr\} \right] \right\}
                \cdot g(|SG(a)|) \cdot h(|J|),
$$                
highlights how the utility function includes
a preference for larger benefit sizes ($m$),
larger subpopulation size ($g$), and parsimonious description ($h$).

Finally, some more comments on the MCMD. 
One possible way to determine $\delta(\cdot)$ is elicitation from physicians. 
We consider here, as an example, using a linear function
\begin{equation}
\delta\bigl\{\ytox (N), \ytox (C) \bigr\} = \delta_0 + \delta_1
\bigl\{\ytox (N) - \ytox (C) \bigr\},
\label{eq:delta}
\end{equation}
as described above.  
To specify the intercept $\delta_0$ and the slope $\delta_1$, one may solicit
(1) an MCMD value in efficacy when assuming no difference in toxicity between $N$ and $C$;
(2) an upper bound of unacceptable difference in toxicity
probabilities between $N$ and $C$ at which $\delta(\cdot)$ takes the
maximum value of expected difference in efficacy between treatment arms, e.g., 0.5 when using the difference in survival probabilities to
compare $N$ and $C$.  Then, we use the MCMD value obtained in (1) to
specify $\delta_0$, and divide (the maximum efficacy value $-$ the MCMD
value) by the upper bound value of unacceptable difference to specify
$\delta_1$.  For example, the MCMD value = 0.2 and the toxicity upper
bound value = 0.2 give $\delta_0$ = 0.2 and $\delta_1$ = $(0.5 - 0.2)$ /
0.2 = 1.5.

Under the outlined framework, we determine an optimal 
rule $a^*$ known as the Bayes rule, which is formally described as   
\begin{equation}
  a^*\ =\arg\max_a \ \ U(a).
\label{decision}
\end{equation}
In addition to $a^*$, we also recommend to report slightly suboptimal choices. 
For example, the top 5 instead of only the formal Bayes rule $a^*$ in \eqref{decision}.
This mitigates undesirable sensitivity with respect to technical
details in the utility function \eqref{Utility} and the assumed 
probability model (related to technical convenience rather than expert judgment).

\section{Probability model}
\label{sec:ProbModel}

\noindent
We model the TTE outcome with an AFT model,  
$y_i = \eta (z_i,{\bx}_i,\bmu) + u_i$, $i = 1, \dots, n$, 
where $\eta (\cdot)$
denotes a mean function (``linear term'' in a traditional AFT model),
$\bmu$ indexes the model for $\eta (\cdot)$ (details below), and $u_i$
is a residual. 
Since the earlier decision of the selection of the subgroup report did
not rely on any details of the probability model, we are free to use
any model to fit the data.  For an optimal fit of the data without
restriction to a parametric family, we propose a non-parametric
Bayesian model.  We use a Bayesian random forest model, BART (Chipman
{\it et al}., 2010; Hill, 2011), for the mean function and assume a
non-parametric P\'olya tree (PT) prior (Lavine, 1992) 
for the residual distribution $G(u_i)$. 
The nonparametric priors avoid strict parametric
assumptions.  Let $\eta_{BT}(z_i,{\bx}_i,\bmu)$ denote the BART mean
function, and let $G\sim \PT(\cA, G_0)$ denote the PT prior for
the residual distribution $G$. 
In summary, 
\begin{align}
& \begin{multlined}[c]
y_i = \eta_{BT} (z_i,{\bx}_i,\bmu) + u_i, \ \ \ i = 1,\dots, n, {}\\ 
u_i \sim G,\, i.i.d., \ \ \ G \sim \PT(\cA,G_0),
\label{AFTmodel}
\end{multlined} 
\end{align}
subject to $G(-\infty,0) = 0.5$.
The first set of hyperparameters, ${\cA}$, defines the probability of
nested sequences of partitions of a sample space via beta distributions.  
The second hyperparameter, $G_0$, defines centering of the random
probabilty measure $G$. 
We assume $G_0 = N(0,\sigma^2)$.  

We use BART as a prior for the mean function 
mainly because of the availability of a very
computation-efficient implementation as public domain R package.  BART
consists of two parts, a sum-of-trees model (random forest) and a
regularization prior on the parameters of the model.
The BART
model allows to naturally incorporate main effects as well as
interactions.  Let $\bmu$ generically denote all BART parameters
including in particular terminal node parameters that represent main
and interaction effects.  
For the residual distribution, we choose a PT prior
over other alternative non-parametric priors
because it naturally lends itself to restricting to zero median as
needed to keep the interpretation as residual distribution (Lavine,
1992).
In addition, a PT prior allows a closed form
expression of the marginal distribution of residuals
$(u_1,\ldots,u_n)$ in \eqref{AFTmodel}, 
that is, 
$$
  G_{mg}(u_1,\ldots,u_n \mid \sigma^2) \equiv
  p(u_1,\ldots,u_n) = \int \prod_{i=1}^n G(u_i) dp(G),
$$
(M\"uller {\it et al.}, 2015). 
The conditioning on $\sigma^2$ in $G_{mg}$ arises
because we assume $N(0,\sigma^2)$ for $G_0$ in the PT prior.   
Semi-parametric AFT models with PT priors have been used before in
Hanson (2006) and Walker and Mallick (1999).  See Ibrahim {\it et al.}
(2001) for a review.

For the probability model for a binary toxicity outcome,  
refer to Morita and M\"uller (2017).

\section{Prediction using MCMC}
\label{sec:MCMC}

\noindent 
We use MCMC posterior simulation  (Gilks {\it et al.}, 1996)
to generate 
posterior Monte Carlo samples of $\bmu$ and $\sigma^2$.
Recall that $y_i\unobs = y_i^o + \kappa_i$ for
censored cases ($\gamma_i = 0$), that is, $\kappa_i$ together with the
censoring time implicitly defines $y_i\unobs$.
Also, let  $\by = (\by\obs, \by\unobs)$ denote the complete data, 
and let 
$\bmeta_{BT} (\bz, \bX, \bmu) =\bigl(\eta_{BT} (z_i,{\bx}_i,\bmu);\;
i=1,\ldots,n \bigr)$. 
In summary, the MCMC proceeds in three steps:
\underline{Step 1,} updating $\bmu$ to obtain new $\bmeta_{BT} (\bz,
\bX, \bmu)$;
\underline{Step 2}, updating $\sigma^2$; and
\underline{Step 3}, obtaining $\by^{\rm
  unobs}$ by imputing $\bkpa$ for censored cases.  
In other words, we iterate over sampling from the
full conditional posterior distributions
$p(\bmu \mid \bz, \bX, \by\obs, \by\unobs, \sigma^2)$, 
$p(\sigma^2 \mid \bz, \bX, \by\obs, \by\unobs, \bmu)$,  
and
$p(\bkpa \mid \bz, \bX, \by^o, \bgamma, \bmu, \sigma^2). $
In Step 3, we implicitely update $y_i\unobs$.
The generated posterior Monte Carlo sample allows us to evaluate 
posterior predictive distributions $p_{pp}(\ty \mid \tbx, \tz, \DTTE)$ as
Monte Carlo averages. 
The scheme integrates the BART method (Chipman {\it et al}., 2010) and
the setup of Walker and Mallick (1999, Sec. 3).   

We add some more implementation details for 
Steps 1 through 2.  
We use Metropolis-Hastings (M-H) type transition probabilities in
\underline{Step 1}, using the BART package
(R Package \texttt{BayesTree}; Core Team: R, 2014)
to generate a proposal, and an
appropriate M-H acceptance probability 
to account for the fact that we replace the normal residual
distribution that is assumed in the BART software
by a PT random probability measure.
To see the right M-H acceptance probability consider 
\eqref{AFTmodel}, with $G$ marginalized out analytically, i.e.,
$(u_1, \ldots, u_n) \sim G_{mg}(u_1,\ldots,u_n \mid \sigma^2)$.
The conditional posterior for $\bmu$ is therefore determined by the
BART prior $p(\bmu)$ and the likelihood $G_{mg}(\bu \mid
\sig^2)$, i.e., 
$$
p(\bmu \mid \bz, \bX, \by, \sigma^2) \propto
p(\bmu)\, G_{mg} (\bu \mid \sigma^2),
\mbox{ with } \bu = (u_1,\ldots,u_n),
   u_i = y_i - \eta_{BT} (z_i,{\bx}_i,\bmu).  
$$
The definition of $u_i$ makes $G_{mg}(\bu \mid \sigma^2)$ implicitly a
function of $\bmu$.  On the other hand, the BART software assumes the
same prior $p(\bmu)$, but normal residuals
$u_i \sim N(0,\sigma^2)$.
Let $q(\bu \mid \bmu, \sig^2)$ denote the normal distribution of
the residuals $u_i$ under the BART model. BART uses the conditional posterior
$q(\bmu \mid \bz, \bX, \by, \sigma^2) \propto p(\bmu)\, q(\bu \mid \bmu, \sig^2)$.
We use $q(\bmu \mid \bz, \bX, \by, \sigma^2)$ as the
proposal distribution for an M-H transition probability that updates
$\bmu$ (and therefore implicitly $\eta_{BT} (z_i,{\bx}_i,\bmu)$).
Importantly note that the above two posteriors differ only by the likelihood.
More specifically, $G_{mg}(\bu \mid \sigma^2)$ 
includes a factor that is identical to the normal likelihood
$q(\bu \mid \bmu, \sig^2)$. This is due to the use of
$G_0=N(0,\sig^2)$ as the centering distribution in the PT prior.
In the end, the remaining factors in $G_{mg}(\cdot)$ are left as
the M-H acceptance probability. 

In \underline{Step 2}, the conditional posterior for $\sig^2$ is determined by
\eqref{AFTmodel}, again with $G$ marginalized out analytically, as 
$$
p(\sigma^2 \mid \bz, \bX, \by, \bmu) \propto p(\sigma^2)\,G_{mg}(\bu \mid \sigma^2).
$$
We assume a scaled inverse gamma prior
$p(\sigma^2)$.  To implement a M-H algorithm to update $\sigma^2$, we
use the proposal posterior distribution $q(\sigma^2 \mid \bz, \bX,
\by, \bmu) $ $\propto p(\sigma^2)\, q(\by \mid \bmu, \bz, \bX,
\sigma^2)$, again with the conjugate normal likelihood $q(\by \mid \bmu,
\bz, \bX, \sigma^2)$ as in Step 1, implying an inverse-gamma posterior
$q(\sig^2 \mid \ldots)$. 
Similarly to Step 1,  the acceptance ratio 
in the M-H transition probability in Step 2 simplifies greatly. 

For a binary toxicity outcome, refer to  
Morita and M\"uller (2017) for details.

\section{Simulation study design}
\label{sec:SimStudy}

\noindent
To evaluate the operating characteristics of the proposed method,
BaPoFi-TTE, we simulated 1,000 hypothetical realizations of the
RCTs with a balanced
design with equal 1:1 allocation to the two treatment arms (new
treatment $N$ vs. control $C$).  We performed the simulation study
with ten covariates ($p = 10$) and a total sample size of $n=400$
under seventeen assumed true scenarios.  For the TTE data, we randomly
selected censored cases to induce 10$\%$ censoring.  
Additional simulations were carried out with
sample sizes $n = 100, 200, 300, 400$, and 500, and a censoring
proportion 50$\%$, still with $p = 10$. 

We will later discuss frequentist summaries under 
repeat simulation.
These summaries, e.g., true positive rate, require the notion of a
``true subpopulation'' under the simulation truth.   
However, the simulation truth only defines true sampling models for the TTE and toxicity outcomes denoted by
$p^{\,0}_{\rm{TTE}} (y \mid z, \bx^{\rm{TTE}}, \bth^{\,0,{\rm{TTE}}})$ and 
$\ptox (\ytox \mid z, \bxtox, \bthtox)$, respectively (see below for details).
Importantly, the specifications of $p^{\,0}_{\rm{TTE}}(\cdot)$ and
$\ptox(\cdot)$ do not include any notion of true subgroups.   
Instead we define a true subgroup as follows.  
Let $U^0(a)$ denote the utility ($\refeq{Utility}$) under the
simulation truth, that is, we replace the calculation in
\eqref{PCTETTE} and the expectation in $\PCTETox(\tbx, \Dtox)$ by
expectations 
under $p^{0}_{\rm{TTE}} (y \mid z, \bx^{\rm{TTE}},
\bth^{0,{\rm{TTE}}})$ and
$\ptox (\ytox \mid z, \bxtox, \bthtox)$.
Under each simulation scenario, we computed $U^0(a)$ for all
subgroups.    
The subgroup with the highest $U^0(a)$ value is the
 ``true subpopulation''  
for the subgroup report.

\subsection{Data generation} 
\label{sec:Dgeneration}

We first simulate TTE outcomes $y_i$ under assumed simulation truths
$p^{0}_{\rm{TTE}} (y \mid z, \bx^{\rm{TTE}}, \bth^{0,{\rm{TTE}}})$
using eleven different scenarios for $p^{0}_{\rm{TTE}}$ with no
difference in toxicity between treatment arms $N$ and $C$
(we added a superscript on $\bx\tte$, to allow for
different sets of baseline covariates for TTE and toxicity models). 
This defines scenarios 0 and E1 through E10. 
Next, we include two scenarios (T1 and T2) for the toxicity outcome, 
with simulation truth 
$\ptox (\ytox \mid z, \bxtox, \bthtox)$.  We then combine the two
toxicity scenarios with three out of the TTE scenarios to assume six
scenarios for the efficacy-toxicity trade-off evaluation
 (scenarios E1*T1 through E4*T2, below). 
For both,
the TTE and toxicity outcomes, we assumed that all covariates were
continuous, and generated $x_{ij}$ independently from a normal {\it
N}(0,1).
We briefly summarize the data generation below. 

\ul{Scenario 0}\; is an overall null case ($H_0$) assuming no difference
in both, the TTE and toxicity outcomes between arms $N$ and $C$.
\ul{Scenario E1} is an overall alternative case ($H_1$) with an overall treatment effect of arm $N$. 
As shown in Figure 1, \ul{Scenarios E2} (1-L), \ul{E3} (1-M), and \ul{E4} (1-S) are scenarios with one predictive covariate ($x_1$) and large
(67\%, E2), moderate (50\%, E3), and small (33\%, E4) subpopulations. 
Here percent are the subgroup size as \% of the entire population. 
\ul{Scenarios E5} (2-M), \ul{E6} (2-S) and \ul{E7} (2-S) are scenarios with two predictive covariates ($x_1$, $x_2$) and moderate (45\%,
E5), small (25\%, E6), and small (22\%, E7) subpopulations. 
\ul{Scenarios E8} (1-S) and \ul{E9} (1-M) assume that an inward subset and distant subsets are sensitive, respectively.
\ul{Scenario E10} (2-M$^{Ls}$) is a case with two predictive
covariates ($x_1$, $x_2$) and an $L$-shaped sensitive subpopulation of
moderate size (55\%).   
For toxicity, \ul{Scenarios T1} (1-S) and \ul{T2} (1-L) are scenarios for small (33\%, T1) and large (67\%, T2) subpopulations.  
Both are defined with one predictive covariate ($x_6$) which is different from those for efficacy ($x_1$, $x_2$). 

Combining scenarios E1, E2, and E4 with T1 and T2, we assume \ul{Scenarios E1$\ast$T1} (1-L) and \ul{E1$\ast$T2} (1-S) with one predictive covariate ($x_6$) and assume \ul{Scenarios E2$\ast$T1} (2-M), \ul{E2$\ast$T2} (2-S), \ul{E4$\ast$T1} (2-S), \ul{E4$\ast$T2} (2-VS: very small, 11\%) with two predictive covariates ($x_1$, $x_6$).

For the TTE outcome, we use the following log-linear model with a linear term $g_{\rm{TTE}}(\cdot)$ to generate data for patient $i$, 
\begin{equation} 
  y_i = g_{\rm{TTE}} (z_i, \bx_i^{\rm{TTE}}, \bth^{0,{\rm{TTE}}}) + e_i, \ \ \  e_i \sim N(0, s^2),
\label{eq_simdata}
\end{equation}
where $\bth^{0,{\rm{TTE}}} = (\beta_0^{0,{\rm{TTE}}},
\bbeta^{0,{\rm{TTE}}}, \gamma^{0,{\rm{TTE}}})$ with
$\beta_0^{0,{\rm{TTE}}}$ for the overall treatment effect of arm $N$,
$\bbeta^{0,{\rm{TTE}}}$ for the covariates, and
$\gamma^{0,{\rm{TTE}}}$ for the interaction effect between treatment
and predictive covariate(s).  That is, we use $g_{\rm{TTE}} (z_i,
\bx_i^{\rm{TTE}}, \bth^{0,{\rm{TTE}}})$ = $\beta_0^{0,{\rm{TTE}}} z_i +
h_{\rm TTE}(\bx_i^{\rm{TTE}}, \bbeta^{0,{\rm{TTE}}}) +
\gamma^{0,{\rm{TTE}}} \ z_i \ I(\cdot)$ and $h_{\rm
TTE}(\bx_i^{\rm{TTE}}, \bbeta^{0,{\rm{TTE}}}) = 0.1x_{1,i} +
0.05x_{2,i} - 0.1x_{3,i} - 0.1x_{4,i} + 0.05x_{5,i}
-0.05x_{1,i}x_{3,i}$.
Here, $I(\cdot)$ denotes an indicator function to specify the
sensitive subpopulation for each scenario explained above.  For
example, Scenarios E2 and E5 use $I(x_{1,i} \ge Q_1^{33})$ and
$I(x_{1,i} \ge Q_1^{67} \cup x_{2,i} \ge Q_2^{33})$, respectively.  As
described in Section \ref{sec:Subgroups}, $Q_j^{q}$ indicates a $q\%$
quantile value of $x_j$.
We set the residual variance $s^2$=$1^2$.  

To generate a binary toxicity outcome for patient $i$ under the
toxicity scenarios T1 and T2, we use a logistic model for 
$p(\ytox=1 \mid \bx\tox, \bth\tox)$ with the linear term $\gtox (z_i, \bxtox_i,
\bthtox) = \htox(\bxtox_i, \bbetatox) + \gammatox \ z_i \ I(x_{6,i}
\le Q_6)$ and use $\htox(\bxtox_i, \bbetatox) = 0.05x_{6,i}
-0.1x_{7,i} - 0.1x_{8,i} + 0.05x_{9,i} + 0.1x_{10,i}
-0.05x_{6,i}x_{8,i}$.  That is, we assume no overall difference in
the probability of toxicity 
between arms $N$ and $C$ (to simplify the
efficacy-toxicity combination scenarios).

Each scenario above  implies  survival
probabilities $S(\tau)$ and overall (marginal) toxicity
probabilities ($Pr({\rm tox})$), for arms $N$ and $C$ in the sensitive
and non-sensitive subpopulations.
We use $\tau$ = 90 (days) for all the efficacy scenarios.
The implied values of the differences in $S(\tau)$ and in $Pr({\rm tox})$
between arms $N$ and $C$ in the sensitive subpopulations used for in
the simulation study are shown in Figures 2a and 2b. 
For the TTE outcome, the simulation truth implies $S(\tau)$=0.20 
in arm $C$ for all efficacy scenarios.
For Scenarios E2 through E10, the simulation truth implies $S(\tau)$=0.30 in arm N, 
that is, $S(\tau)$ of arm $N$ is 10$\%$ higher than that of arm $C$ in
the non-sensitive subpopulation. 
For the toxicity outcome, the simulation truth implies 
$Pr({\rm tox})$=0.10 in arm $C$ and no difference in $Pr({\rm tox})$ is assumed
between arms in the non-sensitive subpopulation for the two scenarios.

\subsection{Frequentist operating characteristics} 
\label{sec:Freqeval}

\noindent Recall that the BaPoFi method reports the top 5 subgroups
with the five highest utilities $U(a)$.  Letting the superscript $^c$
denote the absence of the specific report in the top 5 subgroups, we
evaluated several errors, as frequentist rates.  We use the following
six error rates $\Pr(a \mid b)$, where $a$ ($a^c$) indicates 
that a decision is (is not) among the top 5 decisions 
and $b$ behind the conditioning bar refers to a simulation truth (with a
slight abuse of the conditioning bar).
Here $\Pr(\cdot)$ denotes frequentist probabilities under repeat
simulation.  
Thus, for example,  $\Pr({\it null} \mid H_0)$ 
is the proportion of trials under the simulation truth $H_0$ in which
the top 5 reports contain $a$ = {\it null}. 
We report six summaries that can be  conveniently summarized by true decision rate
(TDR) and false decision rate (FDR). 
\ul{TDR} includes
true negative rate (TNR) = ${\rm Pr}({\it null} \mid H_0)$, 
true positive rate (TPR) = Pr$({\it all} \mid H_1)$, and  
true subgroup rate (TSR) = Pr$(a \mid H_a)$. 
Under a given simulation truth only one of these three rates applies.
\ul{FDR} includes
false negative rate (FNR) = Pr$({\it null} \mbox{ and } a^c \mid H_a)$ 
under $H_a$ or Pr$({\it null} \mbox{ and } {\it all}^c\mid H_1)$ under $H_1$,   
false positive rate (FPR) = Pr$({\it all} \mbox{ and } a^c \mid H_a)$,   and
false subgroup rate (FSR) = 1 $-$ Pr$(a \mbox{ or } {\it null} \mbox{ or }
{\it all} \mid H_a)$ or 1 $-$ Pr$({\it null} \mbox{ or } {\it all} \mid H_1)$.  
FDR is the sum of FNR, FSR and, in the case of a simulation truth
$H_a$, FPR. 
In other words, under $H_a$ and $H_1$, FSR is the proportion of trials which report a false subgroup. 
Note that type I error (T1E) is included as 1$-$TNR = Pr$({\it null}^c \mid H_0)$.   

We use the tuning parameters $\nu, \zeta, u_0$ in the utility function \eqref{Utility} to achieve desired levels of the described frequentist operating characteristics. 
We adjust $u_0$ to achieve desired type-I error rate under a null simulation truth.
In our implementation, we varied $\nu$ and $\zeta$ on a grid between $0.10$ and $0.50$ (in increments of $0.05$). 
After determining $(\nu, \zeta) = (0.25, 0.15)$ under $p = 10$ and $n = 400$, we set $u_0 = -0.304$. 
When using the approach for data from a real clinical study, one may specify the tuning parameters, based on results of the repeat simulation using the actual numbers of covariates and patients and the summaries of efficacy and toxicity outcomes.

\subsection{Simulation results} 
\label{sec:SimResults}

\noindent 
Figure 2a summarizes simulation results with four operating characteristics for $n$ = 400 and $p$ = 10 under the ten efficacy clinical scenarios.
Figure 2b shows those under the six efficacy-toxicity trade-off scenarios.
Under \ul{Scenario 0} (null scenario: $H_0$), the TDR was 0.95, that is, the TIE was controlled at 0.05.

Under \ul{Scenarios E1-E4}, we find high TDRs under 40$\%$ of the difference in survival probabilities, $S(\tau$), between $N$ and $S$, while moderate TDRs under the difference of 30$\%$. 
Taking that the difference of 40$\%$ in $S(\tau$) indicates a substantial efficacy of treatment $N$ into account, the BaPoFi-TTE performs sufficiently well under a regular sample size ($n$=400) for randomized clinical trials.

Under \ul{Scenarios E5-E7} with two predictive covariates, 
BaPoFi-TTE maintains acceptable rates.  
Similarly to the above four scenarios, performance of BaPoFi-TTE improves as the difference in $S(\tau$) increases.
It should be noted that the BaPoFi-TTE performs well regardless of the subpopulation size.
These results further suggest higher flexibility of BaPoFi-TTE in subpopulation finding.

Under \ul{Scenarios E8 and E9} with the challenging inward and distant subsets and under \ul{Scenario E10} with the more challenging $L$-shaped
simulation truth, BaPoFi-TTE continues to work reasonably well as the difference between treatment arms increases to 50$\%$. 
  
\vskip .1in

As shown in Figure 2b, under the clinical scenarios combining the three efficacy and two toxicity scenarios, the BaPoFi-TTE imbedding a utility function accounting for the efficacy-toxicity trade-off shows good performance in population finding.
Overall, the TDR values increase as the differences between treatment arms in efficacy and toxicity outcomes become larger in the respective subgroups, as expected. 
However, several points should be noted as follows.

Under \ul{Scenario E1$\ast$T1} where treatment $N$ has a high efficacy in the entire population but gives a high toxicity occurrence in a small patient population, the BaPoFi-TTE frequently fails to include (the true) $H_a$ among the top five subgroup reports and wrongly includes $H_1$ as the high FPR values indicate.  
This undesirable performance (FPR = 0.52) is also observed under \ul{Scenario E1$\ast$T2} with a large toxic subpopulation when the efficacy difference is 0.40 and the toxicity difference is 0.25. 
Under the other scenarios, in cases the patient subpopulation to be reported has a small size, the BaPoFi-TTE performs reasonably well if sufficient contrasts are observed between the efficacy and toxicity outcomes.  
These findings suggest the importance of pre-analysis discussion with physicians about which toxicities should be accounted and how to define a function $\delta(\cdot)$ and its parameters (i.e., $\delta_0$ and $\delta_1$) for the efficacy-toxicity trade-off in a utility function. 

\vskip .1in
Figure 3 summarizes TDR (=TSR) values under \ul{Scenarios 3, 5, and 7} with $p = 10$ and sample sizes $n = 100, 200, 300, 400$, and $500$ and censoring proportions $= 10$ and $50\%$.
Overall, the BaPoFi-TTE performs better as the sample sizes increases for the three scenarios, as expected. 
This exploratory investigation suggests that when the number of covariates of interests is moderate, say around $p = 10$, a sample size of larger than or equal to 400 may be required to provide around 70\% TDR to find a small to moderate size (say 25 to 50\% of $n$) subpopulation defined by two predictive covariates, while 200 may be enough for one-covariate cases, unless the censoring proportion is not so high. 

The acceptance rates of $\bmu$ and $\sigma^2$ in Steps 1 and 2 in the MCMC computations were on average 0.191 and 0.574, respectively, under Scenario 2. 
Similar values were observed under the other scenarios.

\section{Application to a phase III trial}
\label{sec:Appli}

\noindent We applied the proposed BaPoFi-TTE method to 
the study that we already briefly introduced earlier, in Section \ref{sec:MotivaingEx}.
In this phase III trial, 544 and 546 (in total, 1090) patients
received treatment $N$ and treatment $C$, respectively (Kaufman {\it et al.}, 2015).  
In this application,  
we analyzed OS as the efficacy TTE outcome.
For the toxicity evaluation, we selected the occurrence of grade 4
hematologic toxicities, that is, either of neutropenia, leukopenia,
anemia, and febrile neutropenia, because a noticeable difference in
the grade 4 hematologic toxicities was noted. 

We compared the differences in survival probabilities at
$\tau=720$ days between arms $N$ and $C$.  We additionally
carried out the analysis using two alternative time horizons with
$\tau=690$ and $\tau=750$ to evaluate the sensitivity of the analysis
to the time-cutpoints.  For the efficacy analysis, we set $\delta_0 =
0.2$ and $\delta_1=0.0$ in the MCMD \eqref{eq:delta}, implying a fixed
MCMD value of $\delta_0$.  We evaluate the efficacy-toxicity
trade-offs using $\delta_1=1.5$ and $\delta_0 = 0.2$
 (compare the brief example after \eqref{eq:delta}). 
For each of the efficacy and toxicity
analyses, we selected candidate covariates from the available baseline
patient covariates.  
Noting that predictive and risk factors need not
be the same for efficacy and toxicity outcomes, we
allowed for partially different sets of covariates for the efficacy
and toxicity analyses.  We used $`$age', $`$body mass index (BMI)',
$`$number of prior chemotherapy regimens for advanced/metastatic
disease [0/1/2$\le$]', $`$disease progression within 60 days after the
last dose of chemotherapy [yes/no]', $`$triple negative (TN)
[yes/no]', and $`$race [white/non-white]' in common for the efficacy
and toxicity analyses.  We include $`$site of disease
[visceral/non-visceral only]', $`$adjuvant therapy [yes/no]',
$`$neo-adjuvant therapy [yes/no]' only in the efficacy analysis, while
containing $`$performance status [0/1/2$\le$]' only in the toxicity
analysis.
Patients whose statuses of the human epidermal growth factor receptor
2 (HER2), estrogen receptor, and progesterone receptor are all
negative are classified into $`$TN [yes]'.

From the total 1090 patients, we excluded 13 due to missing data in the covariates of $`$BMI' or $`$site of disease'.
We then analyzed data from 1077 patients ($N$: 538, $C$: 539). 
We implement BaPoFi-TTE using the  
$p=8$ covariates. We set $u_0 = -0.327$ to control T1E at 0.05.
In addition, we determined to use $(\nu, \zeta) = (0.20, 0.15)$ based
on the earlier reported simulation study under $n$=400 and taking the
increase in the total sample size to $n$=1077 into account.   
As an ad-hoc sensitivity analysis, we carried out alternative analysis
using $(\nu, \zeta)$=(0.30,0.15). 

Figure 4 shows the top 5 subgroup reports, in descending order from
the left, using the same format as in Figure 1.   
The first row shows the result from the efficacy analysis using the OS outcome only.
Based on the frequency of the clinical covariates that appear in the
top 5, we note TN as the most probable sensitive covariate
for treatment $N$.  
Four additional covariates, 
$`$site of disease [visceral]', 
$`$disease progression within 60 days [yes]', 
$`$race [white]', and
$`$age' 
are promising candidates to identify more focused sensitive
subpopulations for treatment $N$ in combination with $`$TN [yes]'.   
The second row shows results from the efficacy-toxicity trade-off analysis.
The results of this analysis give us deeper insights for evaluating
possible patient subpopulations being more suitable for treatment
$N$. 
First, 
the subgroup specified by the single covariate TN went down below the top 5.  
Then, the analysis identified a new covariate $`$BMI' with different
cutoffs (66.7\% and 33\% points) to define the first and fourth
best subgroups. 
Comparing the occurrences of grade 4 hematology toxicity between
treatment arms with respect to the BMI subgroups, the difference in
the toxicity occurrence between arms increased as the BMI level
decreased. 
In summary, we report that 
the subgroup of triple negative patients with higher baseline BMI level
is a more desirable subpopulation for treatment $N$ when taking the
efficacy-toxicity balance into account. 

As shown in the bottom two rows, the analysis with larger $\nu$=0.30
tends to rate larger subpopulations
more favorably 
compared to the first and second rows. 
Recall that $\nu$ represents preference for a larger population size.
In addition, the sensitivity analyses with the two additional cutoff timepoints $\tau$ (=690 and 750) overall supported the above findings.

\section{Discussion}
\label{sec:Discussion}

\noindent The proposed BaPoFi-TTE method casts the population finding
problem as a formal Bayesian decision problem, that is, we
separate the 
construction of the assumed statistical inference model and a
description of the possible actions and preferences.  The BaPoFi-TTE
method evaluates efficacy-toxicity trade-offs and uses a
semi-parametric failure time model with a random forest method, BART,
and a P\'olya tree prior to fit TTE data.  However, any other
flexible sampling models could be used.  A decision theoretic approach
often depends on the arbitrary definition and choice of a utility
function and its design parameters.  Thus, we report the top 5
subgroups rather than the optimal subgroup under controlling the type
I error rate to mitigate undesirable sensitivity to technical details
in the utility function.
When clinical investigators need more focused recommendations, it may be vital to take advantage of as much prior medical knowledge as possible.

Some more limitations remain.  The described approach does not deal
with missing (not available) covariates as it may occur in real
clinical data analyses.  But using a model-based approach, it is
easy to extend the current method to impute missing data. 
Finally, determining optimal cutoff points
especially for continuous covariates is an important remaining issue.

\vskip 0.3in

\centerline{\sc Acknowledgements}  
\noindent 
We thank Eisai Co., Ltd for providing us with the clinical data from Study 301.
The dataset was transferred from Eisai Co., Ltd under the data transfer agreement between Eisai Co., Ltd and Kyoto University.
Satoshi Morita's research was partially supported by the Project Promoting Clinical Trials for Development of New Drugs (19lk0201061h0004) from the Japan Agency for Medical Research and Development (AMED).
Peter M\"uller's research was partially supported by
NIH/NCI 2R01CA132897.
Hiroyasu Abe's research was partially supported by Japan Society for the Promotion of Science (JSPS) KAKENHI JP19K13822.
We received no financial support from Eisai Co., Ltd for this research.

\vskip 0.3in

\centerline {\sc References}  
\begin{description} \itemsep=0pt

\item Chipman HA, George EI, McCulloch RE. (1998).
Bayesian CART model search.
{\it The Journal of the American Statistical Association} \textbf {93}: 935-948.

\item Chipman HA, George EI, McCulloch RE. (2010).
BART: Bayesian additive regression trees.
{\it The Annals of Applied Statistics} \textbf{4}: 266-298.

\item Core Team: R. (2014). 
A language and environment for statistical computing. R Foundation for Statistical Computing, Vienna, Austria. http://www.R-project.org/.

\item Foster JC, Taylor JM, Ruberg SJ. (2011).
Subgroup identification from randomized clinical trial data.
{\it Statistics in Medicine} \textbf{30}: 2867-2880.

\item Friedman JH. (1991).
Multivariate adaptive regression splines. 
{\it The Annals of Statistics} \textbf {19}: 1-67.

\item Gilks W, Richardson S, Spiegelhalter D. (1996).
{\it Markov Chain Monte Carlo in Practice}.
Chapman \& Hall: London. 

\item  Graf A, Posch M, Koenig F. (2015).
Adaptive designs for subpopulation analysis optimizing utility functions.
{\it Biometrical Journal} \textbf {57}: 76–89.

\item
Hanson, TE. (2006). 
Inference for mixtures of finite P\'olya tree models. 
{\it The Journal of the American Statistical Association}, \textbf{101}, 1548–1565.

\item Hill JL. (2011).
Bayesian nonparametric modeling for causal inference.
{\it Journal of Computational and Graphical Statistics} \textbf{20}: 217-240.

\item Jones HE, Ohlssen DI, Neuenschwander B, Racine A, Branson M. (2011).
Bayesian models for subgroup analysis in clinical trials.
{\it Clinical Trials} \textbf{8}: 129-143. 

\item Kaufman PA, Awada A, Twelves C, Yelle L, Perez EA, Velikova G, et al. (2015).
Phase III open-label randomized study of eribulin mesylate versus capecitabine in patients with locally advanced or metastatic breast cancer previously treated with an anthracycline and a taxane.
{\it Journal of Clinical Oncology} \textbf{33}: 594-601.

\item Lavine M. (1992).
Some aspects of P\'olya tree distributions for statistical modeling.
{\it The Annals of Statistics} \textbf{20}: 1222-1235.

\item Lipkovich I, Dmitrienko A, Denne J, Enas G. (2011).
Subgroup identification based on differential effect search--a recursive partitioning method for establishing response to treatment in patient subpopulations.
{\it Statistics in Medicine} \textbf{30}: 2601-2621.

\item Morita S, M\"uller P. (2017). 
Bayesian population finding with biomarkers in a randomized clinical trial.
{\it Biometrics} \textbf {73}: 1355-1365.

\item Ondra T, Dmitrienko A, Friede T, Graf A, Miller F, Stallard N, Posch M. (2016). 
Methods for identification and confirmation of targeted subgroups in clinical trials: A systematic review.
{\it Journal Biopharmceutical Statistics} \textbf {26}: 99-119.

\item Rasmussen CE, Williams CKI. (2006).
{\it Gaussian Processes for Machine Learning}.
The MIT Press: Cambridge.  

\item Renfro LA, Mallick H, An MW, Sargent DJ, Mandrekar SJ. (2016).
Clinical trial designs incorporating predictive biomarkers.
{\it Cancer Treatment Reviews} \textbf {43}: 74-82.

\item Robert CP. (2007).
{\it The Bayesian Choice}.
Springer-Verlag: New York. 

\item Rubin DB. (1978).
Bayesian inference for causal effects: The role of randomization. 
{\it The Annals of Statistics} \textbf {6}: 34-58. 

\item Schnell PM, Tang Q, Offen WW, Carlin BP. (2016). 
A Bayesian credible subgroups approach to identifying patient subgroups with positive treatment effects.
{\it Biometrics} \textbf {72}: 1026-1036.

\item Simon N, Simon R. (2018). 
Using Bayesian modeling in frequentist adaptive enrichment designs.
{\it Biostatistics} \textbf {19}: 27-41.

\item Sivaganesan S, M\"uller P, Huang B. (2016).
Subgroup finding via Bayesian additive regression trees. 
{\it Statistics in Medicine} \textbf{36}: 2391-2403.

\item Sparapani R, Logan B, McCulloch R, Laud P. (2016). 
Nonparametric survival analysis using Bayesian additive regression trees (BART).
{\it Statistics in Medicine} \textbf {35}: 2741-2753.

\item Twelves C, Awada A, Cortes J, Yelle L, Velikova G, Olivo M, et al. (2016). 
Subgroup analyses from a phase 3, open-label, randomized study of eribulin mesylate versus capecitabine in pretreated patients with advanced or metastatic breast cancer.
{\it Breast Cancer (Auckl)} \textbf{10}: 77-84.

\item Walker S, Mallick B. (1999).
A Bayesian semiparametric accelerated failure time model.
{\it Biometrics}  \textbf {55}: 477-483.

\item Xu Y, M\"uller P, Tsimberidou AM, Berry D. (2018).
A nonparametric Bayesian basket trial design.
{\it Biometrical Journal} DOI: 10.1002/bimj.20170016: 1-15.

\end{description}


\newpage
\normalsize
\centerline{\sl Figure Legends} 
\vskip .3in

\renewcommand{\baselinestretch}{1.8}\selectfont

\noindent Figure 1. 
Clinical scenarios for the simulation study. 
The new treatment provides a higher efficacy in the shadow area than the rest, that is, the area shows the interaction effect between treatment and covariates  ($x_1, x_2$), under each of Scenarios E2 to E10. 
Scenario 0 is the overall null case ($H_0$), while Scenario E1 is overall alternative case ($H_1$).
The new treatment gives a higher toxicity due to covariate $x_6$ in the shadow area than the rest under each of Scenarios T1 and T2. 
Combinations of E1, E2, and E4 with T1 and T2 define six scenarios for the efficacy-toxicity trade-off evaluation. 

\vskip .15in

\noindent Figures 2a and 2b. Simulation results of the BaPoFi-TTE method (2a for efficacy; 2b for efficacy-toxicity) with ($n$=) 400 patients and ($p$=) 10 covariates, summarized in terms of true decision rate (TDR) and false decision rates (FDR). 
TDR unifies true negative rate (TNR) under Scenario 0, true positive rate (TPR) under Scenario E1, and true subgroup rate (TSR) under the others. 
FDR includes false negative rate (FNR), false positive rate (FPR), and false subgroup rate (FSR). 
The scenarios are characterized by the number of predictive covariates (0, 1, 2), the size (L: large, M: moderate, S: small, VS: very small) and shape (rectangular, distant ($Dt$), $L$-shaped ($Ls$)) of sensitive subpopulation, and the differences in $S(\tau$=90) and $Pr({\rm tox})$ between $N$ and $C$ in the sensitive subpopulation, 
denoted by D-$S(\tau)^S$ and D-$Pr({\rm tox})^S$, respectively. 

\vskip .15in

\noindent Figure 3. 
True decision rates (= true subgroup rates) for the BaPoFi-TTE with the number of covariates $p$ = 10, censoring proportions = 10$\%$ and 50$\%$, and for the total sample size values $n$ = 100, 200, 300, 400, 500, under Scenarios 3, 5, and 7.
Differences in $S(\tau$=90) between $N$ and $C$ in the sensitive subpopulation are all 40\%.

\vskip .15in

\noindent Figure 4. 
Application of the BaPoFi-TTE method to a phase III clinical trial, reporting the top 5 subgroups with $(\nu, \zeta)$ = (0.20,0.15) and (0.30,0.15) from the efficacy and efficacy-toxicity trade-off analyses.
\% values in the boxes indicate the subgroup sizes of the selected subpopulations. 
\lq\lq TN" denotes \lq\lq triple negative", \lq\lq PD in 60 days" means \lq\lq disease progression within 60 days after the last dose of chemotherapy", and \lq\lq BMI" denotes \lq\lq body mass index". 
\lq\lq Y" in the parenthesis denotes \lq\lq yes", and $Q_j^{33}$ and $Q_j^{67}$ are the 33\% and 67\% point cutoff values, respectively.

\newpage
\begin{figure}
  \centering
  \includegraphics[width=15cm]{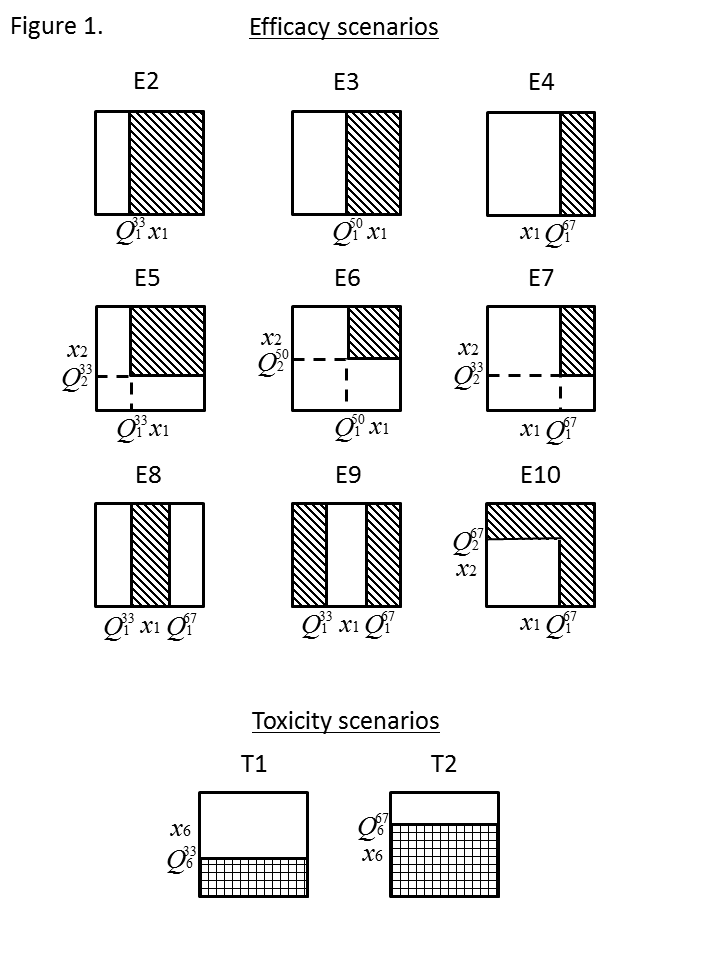}
\end{figure}

\newpage
\begin{figure}
  \centering
  \includegraphics[width=15cm]{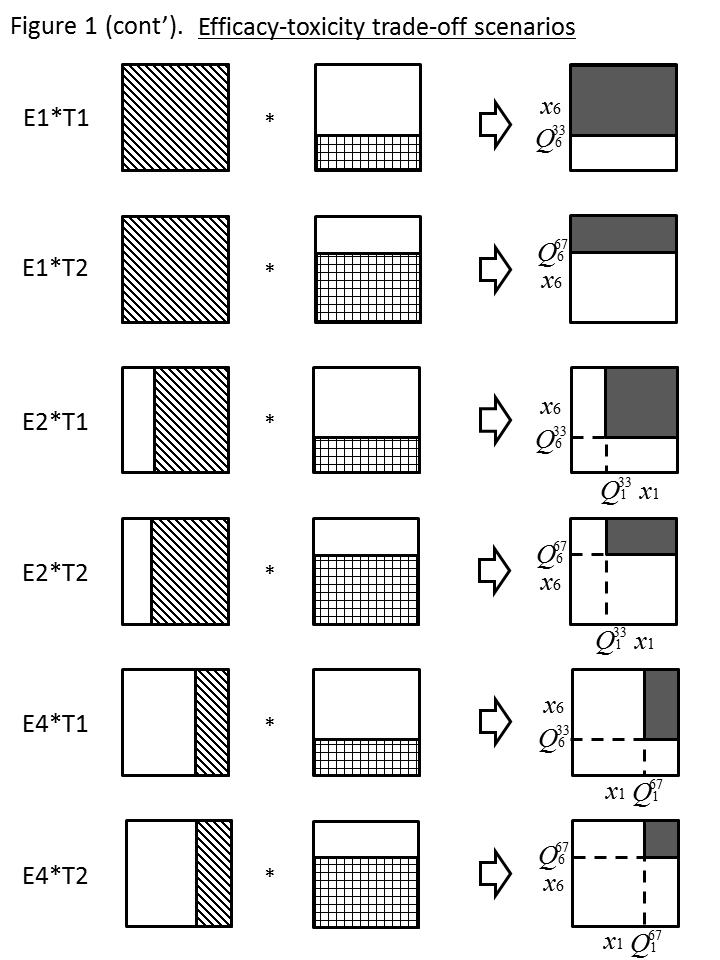}
\end{figure}

\newpage
\begin{figure}
  \centering
  \includegraphics[width=16cm]{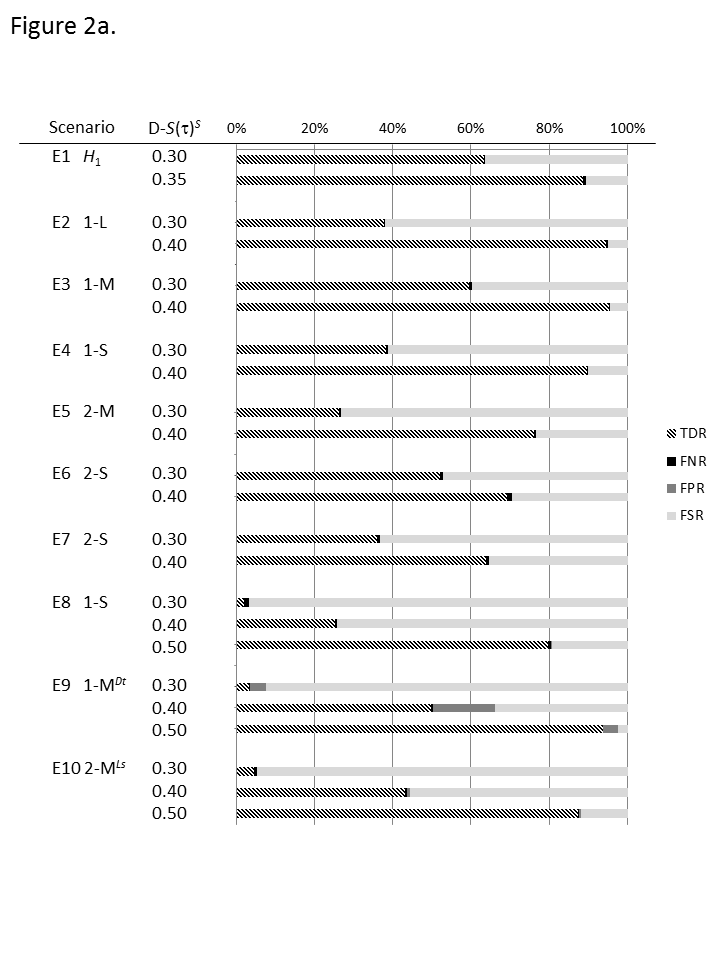}
\end{figure}

\newpage
\begin{figure}
  \centering
  \includegraphics[width=16cm]{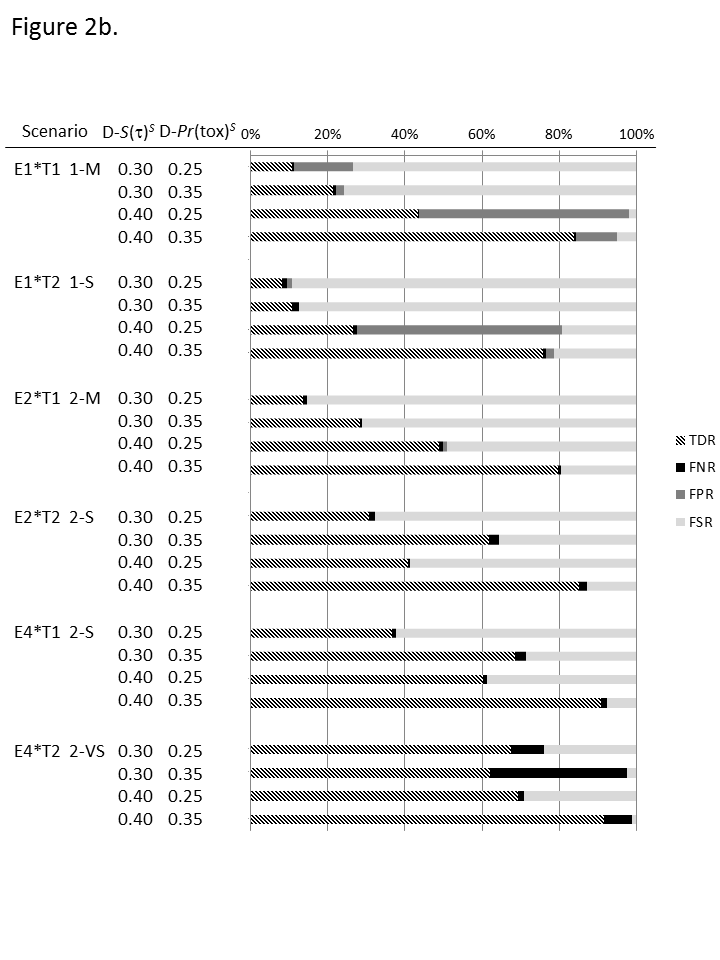}
\end{figure}

\newpage
\begin{figure}
  \centering
  \includegraphics[width=16cm]{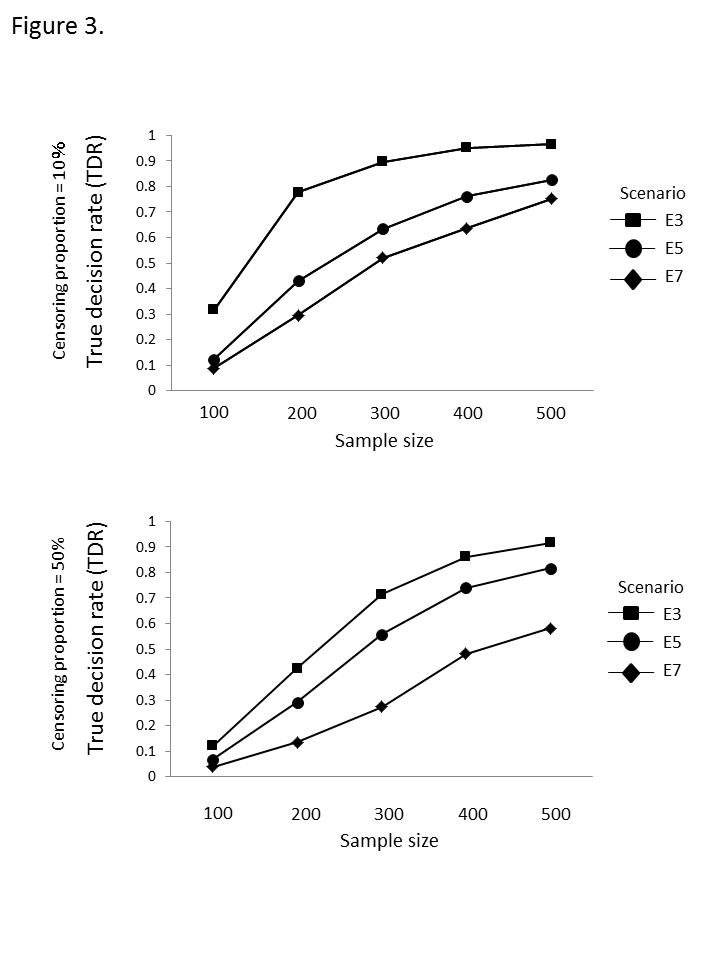}
\end{figure}

\newpage
\begin{figure}
  \centering
  \includegraphics[width=15cm]{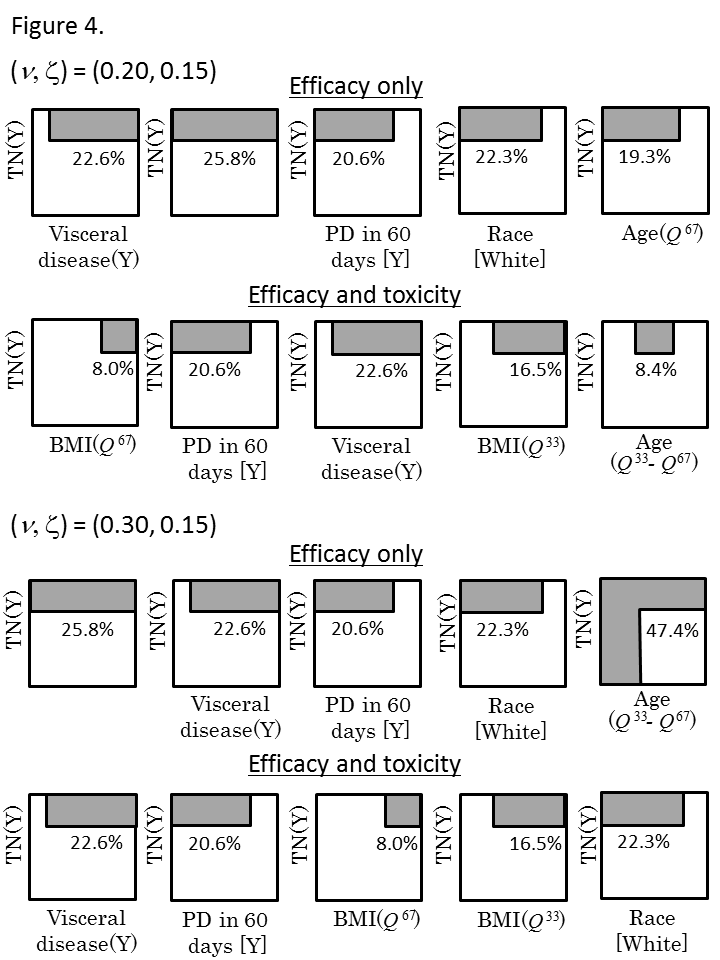}
\end{figure}

\end{document}